\documentclass[aps,amsmath,amssymb,twocolumn]{revtex4}

\usepackage{graphicx}
\usepackage{dcolumn}
\usepackage{bm}
\usepackage{amsmath}
\usepackage{epstopdf}
\usepackage{amsfonts}
\usepackage{amssymb}
\usepackage{epsfig}
\usepackage{tabularx}
\usepackage{color}
\usepackage{soul}
\usepackage[dvipsnames]{xcolor}

\usepackage[colorlinks=true,citecolor=Magenta,urlcolor=Emerald,breaklinks]{hyperref}

\newcommand{\be}{\begin{equation}}
\newcommand{\en}{\end{equation}}
\newcommand{\bea}{\begin{eqnarray}}
\newcommand{\ena}{\end{eqnarray}}

\begin{document}

\title{ISCOs and OSCOs in the presence of a positive cosmological constant in massive gravity}

\author{
\'Angel Rinc\'on {${}^{a}$
\footnote{
\href{mailto:angel.rincon.r@usach.cl}{angel.rincon.r@usach.cl} 
}
} 
Grigoris Panotopoulos  {${}^{b, c}$
\footnote{
\href{mailto:grigorios.panotopoulos@tecnico.ulisboa.pt}{grigorios.panotopoulos@tecnico.ulisboa.pt} 
}
}
Il{\'i}dio Lopes  {${}^{b}$
\footnote{
\href{mailto:ilidio.lopes@tecnico.ulisboa.pt}{ilidio.lopes@tecnico.ulisboa.pt} 
}
}
Norman Cruz  {${}^{a}$
\footnote{
\href{mailto:norman.cruz@usach.cl}{norman.cruz@usach.cl} 
}
}
}

\address{
${}^a$ Departamento de F\'isica, Universidad de Santiago de Chile, Avenida Ecuador 3493, Estaci\'on Central, 9170124, Santiago, Chile.
\\
${}^b$ Centro de Astrof{\'i}sica e Gravita{\c c}{\~a}o-CENTRA, Instituto Superior T{\'e}cnico-IST, Universidade de Lisboa-UL, Av. Rovisco Pais, 1049-001 Lisboa, Portugal.  
\\
${}^c$ Departamento de Ciencias F\'isicas, Universidad de la Frontera, Avenida Francisco Salazar 01145, Temuco - Chile. 
}

\begin{abstract}
We study the impact of a non-vanishing (positive) cosmological constant on the innermost and outermost stable circular orbits (ISCOs and OSCOs, respectively) within massive gravity in four dimensions. The gravitational field generated by a point-like object within this theory is known, generalizing the usual Schwarzschild--de Sitter geometry of General Relativity. In the non-relativistic limit, the gravitational potential differs by the one corresponding to the Schwarzschild--de Sitter geometry by a term that is linear in the radial coordinate with some prefactor $\gamma$, which is the only free parameter. 
Starting from the geodesic equations for massive test particles and the corresponding effective potential, we obtain a polynomial of fifth order that allows us to compute the innermost and outermost stable circular orbits. Next, we numerically compute  the real and positive roots of the polynomial for several different structures (from the hydrogen atom to stars and globular clusters to galaxies and galaxy clusters) considering three distinct values of the parameter $\gamma$, determined using physical considerations, such as galaxy rotation curves and orbital precession. Similarly to the Kottler spacetime, both ISCOs and OSCOs appear. Their astrophysical relevance as well as the comparison with the Kottler spacetime are briefly discussed. 
\end{abstract}

\maketitle

\section{Introduction\label{sec1}}

Current observational data in astrophysics and cosmology indicate that the present Universe is dominated by dark matter and dark energy~\cite{turner}, the~origin and nature of which still remain a mystery. The~dark sector comprises one of the major challenges in modern theoretical cosmology. A~positive cosmological constant, $\Lambda$, \cite{carroll} is the simplest and most economical way to explain the current cosmic acceleration, while, in the past, galaxy rotation curves  provided some of the first and strongest evidence in favor of dark matter~\cite{rubin}. 

Einstein's General Relativity (GR) \cite{einstein} may be extended in several different ways, either in four or in higher dimensions--for instance, the $f(R)$ theories of gravity~\cite{mod1,mod2}, the Brans--Dicke~\cite{Brans:1961sx,Brans:1962zz,Dicke:1961gz} and more generically scalar-tensor theories of gravity in four dimensions, brane models~\cite{brane1,brane2}, and Lovelock theory~\cite{Lovelock:1971yv} in higher dimensional spacetimes. 
In four dimensions, the Einstein tensor is the only second-rank tensor with the following properties: (i) it is symmetric, (ii) it is divergence free, (iii) it depends only on the metric
and its first and second derivatives, and~(iv) it is linear in second derivatives of the metric. 

However, in~higher dimensions, Lovelock's theorem
states that more complicated tensors with the above
properties exist.
Of particular interest is massive gravity~\cite{deRham:2010ik,deRham:2010kj} in~which a static, spherically symmetric solution to the vacuum field equations exists~\cite{Ghosh:2015cva}, generalizing the well-known Schwarzschild solution~\cite{SBH} of GR, and~which is characterized by two new scales, $\Lambda$ and $\gamma$. The~former is relevant for the current acceleration of the Universe, while the latter may be explain the galaxy rotation curves provided that $\gamma \sim 10^{-28}~\text{m}^{-1}$ \cite{Panpanich:2018cxo}.

The impact of a non-vanishing cosmological constant on black hole physics has been extensively investigated over the decades in~an effort to determine
whether or not new effects appear~\cite{Ashtekar:2017dlf}. Certainly, such a study has been extended to other topics into General Relativity, astrophysics, and cosmology.
In particular, as~was recently pointed out by M.~Visser and collaborators~\cite{Boonserm:2019nqq},  new features emerge, such as outermost stable circular orbits (OSCOs), when a positive cosmological constant (no matter how small) is taken into account.

In this respect, OSCOs has been investigated in alternative contexts, for~instance: (i)~accretion disks~\cite{Rezzolla:2003re,Stuchlik:2008dv}, (ii) galaxies~\cite{Stuchlik:2011zz,Sarkar:2014cca}, and  in (iii) modified theories of gravity~\cite{Perez:2012bx,Lee:2017fbq}.
There is a vast literature where models in the context of extended theories of gravity have been studied. To~name a few, within~the context of scalar-tensor theories of gravitation, the~Brans--Dicke theory is considered one of the most natural extensions of General Relativity~\cite{Brans:1961sx,Brans:1962zz,Dicke:1961gz}. 

Based on similar ideas, scale-dependent gravity is an alternative approach, where the coupling constants of the theory are allowed to vary~\cite{SD1,SD0,SD2,SD3,SD4,SD5,SD6,SD7,SD8,SD9,SD10,SD14,SD15,SD16,astro1,astro2,cosmo1,cosmo2}.
In addition to that, in~higher dimensions, another possibility is the well-known Gauss–Bonnet gravity~\cite{Cai:2001dz}, and,~more generically,  Lovelock gravity~\cite{Lovelock:1971yv} in~which higher order curvature corrections are natural.

In this work, our goal is twofold: First, we shall use the recent data reported by the GRAVITY Collaboration to place limits on the parameters of massive gravity. Next, assuming those values, we shall investigate the existence and nature of the stable circular orbits of several different structures in the Universe from~the atomic level to  clusters of galaxies.
In this paper, we will investigate whether OSCOs are also present in light in massive gravity
~\cite{deRham:2010kj,Panpanich:2018cxo} taking into account real values of the massive gravity parameter $\gamma$.

Our work is organized as follows: In the next section, we review the field equations and the vacuum solution of massive gravity. In~the third section, we obtain the allowed range of the scale $\gamma$ using data on the periastron advance of the planet Mercury in our Solar System as well as of the star $S_2$ orbiting around the supermassive black hole at the Galactic center. In Section~\ref{sec4}, we discuss the geodesic equations and the effective potential for test massive particles, while, in the fifth section, we compute the ISCOs and OSCOs of several different structures in the Universe. 

Finally, we summarize our work in the last section with some concluding remarks. We adopt the mostly negative metric signature $(+,-,-,-)$, and~we work mostly in geometrized units where we set the speed of sound in a vacuum as well as Newton's constant to unity, $G=1=c$.

\section{Field Equations and Vacuum Solution in Massive~Gravity\label{sec2}}

We will start by considering the theory of dRGT massive gravity, defined by the action~\cite{deRham:2010kj,deRham:2010ik}, and~we closely follow~\cite{Panpanich:2018cxo}
\begin{equation}
S[g_{\mu\nu}, f_{\mu \nu}] =  \frac{M_{\rm Pl}^2}{2}   \int d^4 x \sqrt{-g} \Bigl[R + m_g^2 \mathcal{U}(g,f) \Bigl] + S_m \,,
\end{equation}
where $S_m$ is the part of the action coming from the matter content, and~we  use the conventional definitions, i.e.,
(i) $M_{\rm Pl}$ is the reduced Planck mass, 
(ii) $R$ is the Ricci scalar, 
(iii) $g$ is the determinant of the metric tensor $g_{\mu \nu}$,
(iv) $m_g$ is the graviton mass, and~finally 
(v) $\mathcal{U}$ is the self-interacting potential of the gravitons. 
In~order to avoid the Boulware--Deser ghost, the~self interactions $U(g,f)$ must be split as follows
\begin{align*}
&\mathcal{U} \equiv \mathcal{U}_2 + \alpha_3 \mathcal{U}_3 + \alpha_4 \mathcal{U}_4 \,, \\
&\mathcal{U}_2 \equiv  [\mathcal{K}]^2 - [\mathcal{K}^2] \,, \\
&\mathcal{U}_3 \equiv [\mathcal{K}]^3 - 3 [\mathcal{K}][\mathcal{K}^2] + 2 [\mathcal{K}^3] \,, \\
&\mathcal{U}_4 \equiv [\mathcal{K}]^4 - 6 [\mathcal{K}]^2 [\mathcal{K}^2] + 3[\mathcal{K}^2]^2 + 8 [\mathcal{K}][\mathcal{K}^3] - 6 [\mathcal{K}^4] \,,
\end{align*}%
where the tensor $\mathcal{K}^{\mu}_{\nu}$ is, then,
\vspace{12pt}
\begin{equation}
\mathcal{K}^{\mu}_{\nu} \equiv \delta^{\mu}_{\nu} - \sqrt{g^{\mu\lambda} \partial_{\lambda} \varphi^a \partial_{\nu} \varphi^b f_{ab}} \,,
\end{equation}
and $[\mathcal{K}] = \mathcal{K}^{\mu}_{\mu}$ and $(\mathcal{K}^i)^{\mu}_{\nu} = \mathcal{K}^{\mu}_{\rho_1} \mathcal{K}^{\rho_1}_{\rho_2} ... \mathcal{K}^{\rho_i}_{\nu}$. 
At this point, we have two different metrics: 
(i)  the  physical metric, $g_{\mu\nu}$, and~
(ii) the fiducial metric, $f_{\mu\nu}$. In~addition, $\varphi^a$ are the St$\Ddot{\rm u}$ckelberg fields. In~what follows, we use the unitary gauge, $\varphi^a = x^{\mu} \delta^a_{\mu}$, thus
\begin{align*}
\sqrt{g^{\mu\lambda} \partial_{\lambda} \varphi^a \partial_{\nu} \varphi^b f_{ab}} &= \sqrt{g^{\mu\lambda} f_{\lambda\nu}} \,.
\end{align*}
 
The gravitational field equations are obtained taking the variation with respect to $g^{\mu\nu}$, and~they are found to be~\cite{Panpanich:2018cxo}
\begin{equation}
G^{\mu}_{\nu} + m^2_g X^{\mu}_{\nu} = 8 \pi G T^{\mu (m)}_{\nu} \,
\end{equation}
where $T^{\mu (m)}_{\nu}$ is the corresponding energy--momentum tensor obtained from the matter Lagrangian. The~massive graviton tensor~\cite{Berezhiani:2011mt,Ghosh:2015cva}, labeled as $X^{\mu}_{\nu}$, is given by
\begin{align}
X^{\mu}_{\nu} &= \mathcal{K}^{\mu}_{\nu} - [\mathcal{K}] \delta^{\mu}_{\nu} - \alpha \left[(\mathcal{K}^2)^{\mu}_{\nu} - [\mathcal{K}]\mathcal{K}^{\mu}_{\nu} +\frac{1}{2}\delta^{\mu}_{\nu} ([\mathcal{K}]^2 - [\mathcal{K}^2])\right] \nonumber \\
&~~ + 3 \beta \left[(\mathcal{K}^3)^{\mu}_{\nu} - [\mathcal{K}](\mathcal{K}^2)^{\mu}_{\nu} +\frac{1}{2}\mathcal{K}^{\mu}_{\nu} ([\mathcal{K}]^2 - [\mathcal{K}^2])\right. \nonumber \\
&~~ \left. - \frac{1}{6} \delta^{\mu}_{\nu} ([\mathcal{K}]^3 - 3 [\mathcal{K}][\mathcal{K}^2] + 2[\mathcal{K}^3]) \right] \,,
\end{align}
where we can redefine the parameters as follows:
\begin{align}
\alpha_3 &= \frac{\alpha - 1}{3} \\
\alpha_4 &= \frac{\beta}{4} + \frac{1 - \alpha}{12}
\end{align}
 
The terms of order $\mathcal{O}(\mathcal{K}^4)$ disappear when taking into account the fiducial metric ansatz~\cite{Panpanich:2018cxo}:
\begin{equation}
f_{\mu\nu} = 
  \begin{pmatrix}
    0 & 0 & 0 & 0 \\
    0 & 0 & 0 & 0 \\
    0 & 0 & C^2 & 0 \\
    0 & 0 & 0 & C^2 \sin^2 {\theta}
  \end{pmatrix} \,,
\end{equation} 
where $C$ is a positive constant.
It is known that the massive gravitons can be treated as an effective fluid where density, $\rho_{g}$, and~pressures, $\{P^{r}_{g}, P^{\theta,\phi}_{g}\}$, depend on the radial coordinate $r$ only.  
The pressures are generically anisotropic with $P^{r}_{g}\neq P^{\theta,\phi}_{g}$, and thus there is a stress generated by the massive gravitons, as~was indicated in Refs.~\cite{Burikham:2016cwz,Kareeso:2018xum}.

Within massive gravity, static, spherically symmetric black hole solutions with mass $M$ in Schwarzschild-like coordinates $(t,r,\theta,\phi)$ are given by the following line element
\begin{equation}
ds^2 = A(r) dt^2 - A(r)^{-1} dr^2 -  r^2 ( d \theta^2 + \sin^2\theta d \phi^2 )
\end{equation}
where the corresponding lapse function, $A(r)$, is found to be~\cite{Panpanich:2018cxo} :
\begin{align}
A(r) &=1 - \frac{2 M}{r} - \frac{1}{3} \Lambda  r^2
+ \gamma  r + \eta 
\end{align}
and where $\Lambda$ acts like a cosmological constant, while the set $\{ \gamma, \eta \}$ are two new parameters coming from massive gravity, which are computed in terms of the graviton 
mass, $m_g$, and~the other parameters of the theory as follows~\cite{Panpanich:2018cxo}
\begin{align}
\Lambda &=  -3m_g^2(1+\alpha+\beta)
\\
\gamma &= -m_g^2C(1+2\alpha+3\beta)  
\\
\eta &= m_g^2C^2(\alpha+3\beta)  
\end{align}

Clearly, when $m_g$ is taken to be zero, the~solution reduces to the usual Schwarzschild geometry of General Relativity. The corresponding metric $A(r)$ has been obtained, for~instance, in~\cite{Ghosh:2015cva,Boonserm:2019mon}. It is important to mention that the strong coupling scale of the dRGT massive gravity theory was estimated in \cite{Boonserm:2019mon}.

Following~\cite{Panpanich:2018cxo}, to~obtain flat space with $\eta=0$, we impose the following condition on $\alpha, \beta$
\begin{align}
     \alpha+3\beta &=0.
\end{align}
 
To~guarantee that the cosmological constant is positive and tiny, in~the following, we  set
\begin{align}
     \beta - \frac{1}{2} &= \zeta
\end{align}
with $\zeta > 0$ being a very small number. It is now easy to verify that $\Lambda$ plays the role of a positive and tiny cosmological constant,
$\Lambda \equiv \frac{3}{l^2}$, with~$l$ being a length~scale.

In the non-relativistic limit, the following relation holds~\cite{landau,Wald:1984rg}:
\begin{align}
2 \Phi(r) + 1 = g_{00}(r) = 1-\frac{2M}{r}-\frac{\Lambda r^2}{3} - 2a_0 r
\end{align}
with $\Phi(r)$ being the gravitational potential, and~we  set $\gamma = -2 a_0$. Thus, in~this modified 
theory of gravity, the~total gravitational potential consists of three terms
\begin{align}
\Phi(r) &= \Phi(r)_N  - \frac{1}{6}\Lambda r^2 - a_0 r
\end{align}
and the gravitational potential energy, $V$, is simply given by $V(r) = m \Phi(r)$, with~$m$ being the mass of a test particle in the fixed gravitational background. Therefore, there are two perturbing potentials, namely one due to the cosmological constant term, and~another due to the linear term in $r$.

Before we continue with our discussion, a~comment is in order here. Within~GR, the Birkhoff theorem ensures that the only static, spherically symmetric solution in empty space is given by the Schwarzschild geometry. Within~massive gravity, however, contrary to GR, the~theorem does not hold~\cite{Jafari:2017ypl}, and~consequently more than one class of solutions may be obtained~\cite{Jafari:2017ypl,Li:2016fbf,Koyama:2011xz,Koyama:2011yg}. This implies that the gravitational field generated by extended mass distributions depends on the shape of the distribution. 

This is an interesting and, at the same time, tricky issue, which requires a very careful examination. In~the present work, however, we can imagine that we restrict ourselves to some class of certain finite mass distributions for which the solution considered here always holds. Therefore, in~the discussion to follow, we  assume that, for all the structures shown in \mbox{Tables \ref{numerical_table} and \ref{numerical_table_II}}  below, the~gravitational field outside the distribution is described by the solution considered in this~work.


\begin{table*}[ht!]
\centering
\begin{tabular}{| c | c | c | c | c | c | c | c | c |} 
\hline
    Object & $M$  & $r^{\text{Kottler}}_{\text{OSCO}}$   &    $r^{\gamma_1}_{\text{OSCO}}$ & $ r^{\gamma_1}_{\text{ISCO}}/6M$ 
                         & $r^{\gamma_2}_{\text{OSCO}}$  & $ r^{\gamma_2}_{\text{ISCO}}/6M$ 
                         & $r^{\gamma_3}_{\text{OSCO}}$ & $ r^{\gamma_3}_{\text{ISCO}}/6M$  
    \\ [0.5ex] 
       \hline
    \hline
   Hydrogen atom  & $ 4.03 \times 10^{-71} $  
   & $6.31 \times 10^{-18}$
   & $2.89 \times 10^{7}$ & $1.00$ 
   & $2.42 \times 10^{6}$  &  $1.00$
   & $2.28 \times 10^{-29}$  & $1.00$
    \\ [0.5ex] 
   \hline
   Earth & $ 1.44 \times 10^{-19} $ 
   & $9.65 \times 10^{-1}$
   & $2.89 \times 10^{7}$ & $1.00$ 
   & $2.42 \times 10^{6}$  &  $1.00$
   & $1.36 \times 10^{-3}$  & $1.00$
   \\
   \hline
   Sun & $ 4.79 \times 10^{-14} $ 
   & $6.69 \times 10$
   & $2.89 \times 10^{7}$ & $1.00$ 
   & $2.42 \times 10^{6}$  &  $1.00$
   & $7.86 \times 10^{-1}$  & $1.00$
    \\
   \hline
   Stellar association & $ 4.79 \times 10^{-13} $ 
   & $1.44 \times 10^{2}$
   & $2.89 \times 10^{7}$ & $1.00$ 
   & $2.42 \times 10^{6}$  &  $1.00$
   & $2.49                     $  & $1.00$
   \\
   \hline
  Open stellar cluster & $ 4.79 \times 10^{-12} $ 
   & $3.10 \times 10^{2}$
   & $2.89 \times 10^{7}$ & $1.00$ 
   & $2.42 \times 10^{6}$  &  $1.00$
   & $7.86                     $  & $1.00$
  \\
   \hline
   Globular cluster & $ 4.79 \times 10^{-9} $
   & $3.10 \times 10^{3}$
   & $2.89 \times 10^{7}$ & $1.00$ 
   & $2.42 \times 10^{6}$  &  $1.00$
   & $2.49 \times 10^{2}$  & $1.00$
   \\
   \hline
  Saggitarius A* & $ 2.06 \times 10^{-7} $ 
   & $1.09 \times 10^{4}$
   & $2.89 \times 10^{7}$ & $1.00$ 
   & $2.42 \times 10^{6}$  &  $1.00$
   & $1.63 \times 10^{3}$  & $1.00$
  \\
   \hline
    Dwarf galaxies & $ 4.79 \times 10^{-5} $ 
   & $6.69 \times 10^{4}$
   & $2.89 \times 10^{7}$ & $1.00$ 
   & $2.42 \times 10^{6}$  &  $1.00$
   & $2.43 \times 10^{4}$  & $1.00$
    \\
   \hline
    Spiral galaxies & $ 4.79 \times 10^{-2} $ 
   & $6.69 \times 10^{5}$
   & $2.89 \times 10^{7}$ & $1.00$ 
   & $2.47 \times 10^{6}$  &  $1.00$
   & $5.40 \times 10^{5}$  & $1.00$
    \\
   \hline
    Galaxy clusters & $ 4.79 \times 10^{1}$ 
   & $6.69 \times 10^{6}$
   & $2.93 \times 10^{7}$ & $1.00$ 
   & $7.60 \times 10^{6}$  &  $1.00$
   & $6.53 \times 10^{6}$  & $1.00$
     \\
   \hline
\end{tabular}
\caption{
OSCOs and ISCOs as a function of mass (in parsecs). We take $(l= 5 \ \text{Gpc})$ for three different values of the parameter $\gamma$. Thus, we have: 
(i) $\gamma_1 = 3.09 \times 10^{-12} \ \text{pc}^{-1}$,
(ii) $\gamma_2 = 2.58 \times 10^{-13} \ \text{pc}^{-1}$ 
and
(iii) $\gamma_3 = -5.16 \times 10^{-14} \ \text{pc}^{-1}$.
\label{numerical_table}
}
\end{table*}


\begin{table*}[ht!]
\centering
\begin{tabular}{| c | c | c | c |} 
\hline
    Object & $M$  & $r^{\gamma_3}_{\text{OSCO}}$ & Astrophysical Relevance?  
    \\ [0.5ex] 
       \hline
    \hline
   Hydrogen atom  & $ 4.03 \times 10^{-71} $  
    & $2.28 \times 10^{-29}$  & Subatomic scales
    \\ [0.5ex] 
   \hline
   Earth & $ 1.44 \times 10^{-19} $ 
   & $1.36 \times 10^{-3}$  & Size of Solar System
   \\
   \hline
   Sun & $ 4.79 \times 10^{-14} $ 
   & $7.86 \times 10^{-1}$  & Rogue planets
    \\
   \hline
   Stellar association & $ 4.79 \times 10^{-13} $ 
   & $2.49                     $  & Rogue planets
   \\
   \hline
  Open stellar cluster & $ 4.79 \times 10^{-12} $ 
   & $7.86                     $  & Size of most globular clusters
  \\
   \hline
   Globular cluster & $ 4.79 \times 10^{-9} $
   & $2.49 \times 10^{2}$  & Open cluster spacing
   \\
   \hline
  Saggitarius A* & $ 2.06 \times 10^{-7} $ 
   & $1.63 \times 10^{3}$  & Globular cluster spacing
  \\
   \hline
    Dwarf galaxies & $ 4.79 \times 10^{-5} $ 
    & $2.43 \times 10^{4}$  & Size of galaxy
    \\
   \hline
    Spiral galaxies & $ 4.79 \times 10^{-2} $ 
   & $5.40 \times 10^{5}$  & Inter-galactic spacing
    \\
   \hline
    Galaxy clusters & $ 4.79 \times 10^{1}$ 
   & $6.53 \times 10^{6}$  & Size of galaxy cluster
     \\
   \hline
\end{tabular}
\caption{
OSCOs  as a function of mass (in parsecs) for $\gamma_3 = -5.16 \times 10^{-14} \ \text{pc}^{-1}$. We take $(l= 5 \ \text{Gpc})$, and~link the corresponding $r_{\text{OSCO}}$ with typical astrophysical scales. 
\label{numerical_table_II}
}
\end{table*}

\section{Periastron Advance in Massive~Gravity\label{sec3}}

In this section, we present the first part of the analysis performed in the present work, namely how to constrain the parameter $\gamma$ using observational data coming from the periastron advance of the planet Mercury around the Sun as well as the $S_2$ star around  Saggitarius $A^*$.

A generic and useful expression for the periastron advance, $\Delta \theta_p$, due to any perturbative potential energy, $V(r)$, beyond~the Newtonian one, is found to be (setting $G=1$)~\cite{Adkins:2007et}
\begin{align}
\Delta \theta_p = \frac{-2L}{Mme} \int_{-1}^{+1} \frac{dz \  z}{\sqrt{1-z^2}}\frac{dV(z)}{dz}
\end{align}
where $L = a (1-e^2)$, the~perturbing potential energy is evaluated at $r=L/(1+e z)$, and~$e$, and $a$ are the eccentricity and 
the semi-major axis of the orbit, respectively.

Let us mention that, as~was indicated in \cite{Zakharov:2018omt}, the~above expression is still valid in modified theories of gravity. The~study of the motion of test particles in a given gravitational background (geodesic equations via the Christoffel symbols) remains the same in all metric theories of gravity, irrespective of the underlying theory. The~general expression for the precession angle in terms of the perturbing potential has been derived considering the orbit $u(\theta)$, where $u=1/r$, and~this expression does not depend on the underlying theory of gravity.

In the present work, clearly there are two contributions beyond the Newtonian potential, namely (i) the cosmological 
constant ($\Delta \theta_p(\text{CC})$), and (ii) the linear term coming from massive gravity ($\Delta \theta_p(\text{MG})$)
as well as the contribution from General Relativity ($\Delta \theta_p(\text{GR})$), and~they are computed 
to be (setting $G=1=c$) \cite{Adkins:2007et}
\begin{align}
\Delta \theta_p(\text{GR}) &= \frac{6 \pi M}{a (1-e^2)}
\\
\Delta \theta_p(\text{CC}) &= \frac{3 \pi a^3}{Ml^2} \sqrt{1-e^2}
\\
\Delta \theta_p(\text{MG}) &= \frac{2 \pi a^2 a_0}{M}\sqrt{1-e^2}
\end{align}
where the total contribution is 
\begin{align}
\Delta \theta_p \equiv \Delta \theta_p(\text{GR})
+
\Delta \theta_p(\text{CC}) + \Delta \theta_p(\text{MG}).
\end{align}
 
From the observational point of view, here, we shall use the precession angle of the planet Mercury~\cite{Obs1,Obs2}
\begin{equation}
\Delta \theta_p - \Delta \theta_p(\text{GR}) = 
 (-0.002 \pm 0.003)'' \ \textrm{per century}
\end{equation}
as well as the precession angle of the $S_2$ star around Saggitarius $A^*$ \cite{Obs1,GRAVITY}
\begin{equation}
f \equiv \frac{\Delta \theta_p}{\Delta \theta_p(\text{GR})} = 1.10 \pm 0.19.
\end{equation}
%
%
 
Finally, regarding the details of the orbit, in~the case of Mercury, we use the following numerical values~\cite{Obs1}
\begin{eqnarray}
M & = & 1.99 \times 10^{30}~\text{kg} \\
a & = & 5.79 \times 10^{7}~\text{km} \\
e & = & 0.20563
\end{eqnarray}
while, in the case of the $S_2$ star, we use the following numerical values~\cite{Obs1}
\begin{eqnarray}
M & = & 4.261 \times 10^{6}~M_{\odot} \\
a & = & 1.54 \times 10^{14}~\text{m} \\
e & = & 0.884649.
\end{eqnarray}

\begin{figure*}[ht!]
\includegraphics[width=0.48\textwidth]{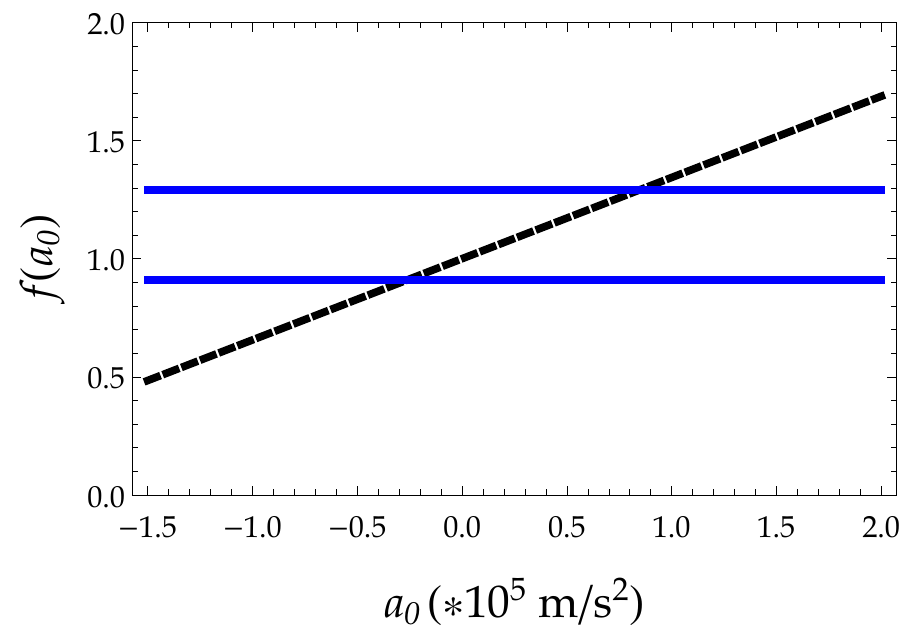} 
\includegraphics[width=0.48\textwidth]{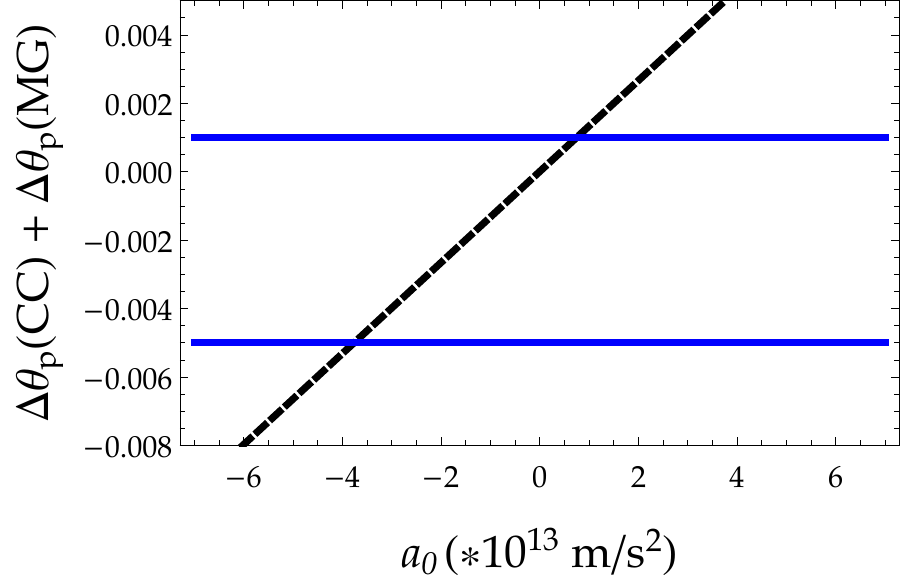} 
\caption{Precession angle against $a_0$ assuming $l=5$ Gpc.
(i) {\textbf{Left Panel:}} Dimensionless ratio $f(a_0)$ for $S_2$ data and its bounds.
(ii) {\textbf{Right Panel:}} Precession angle (deviation from GR) $\Delta \theta_p(\text{CC})$ + $\Delta \theta_p(\text{MG})$ for Mercury data and its bounds.\label{fig:potentialscase1}}
\end{figure*}

In the two panels of Figure~\ref{fig:potentialscase1}, we show, both for Mercury and for the $S_2$ star, the prediction of the theory for the periastron advance as a function of $a_0$ as well as the corresponding observational 
strip. An~allowed window from a lower (negative) to an upper bound (positive) for $a_0$ is obtained. This is the first main result of the present work. The~$a_0=0$, corresponding to GR with a non-vanishing cosmological constant, is included as expected. The~strongest limits come from Mercury, and~those are the ones we shall be using in the discussion to~follow.

The bound on $a_0$ induces a corresponding bound on $\gamma$, which is computed to be
\begin{equation}
-5.16 \times 10^{-14} \ \text{pc}^{-1} \leq \gamma \leq  2.58 \times 10^{-13} \ \text{pc}^{-1}.
\end{equation}
Previously, an~analysis based on galaxy rotation curves showed that, within the dRGT massive gravity, $\gamma \sim 10^{-28}~\text{m}^{-1}$ \cite{Panpanich:2018cxo}.
Finally, we also report on the constraint obtained here using the orbital precession of the $S_2$ star around Saggitarius $A^*$, since, to the best of our knowledge, this is the first attempt to constrain $\gamma$, or~equivalently $a_0$, upon~comparison to the results of GRAVITY Collaboration. We find
\begin{equation}
-2.62 \times 10^{-6} ~\text{m/s}^2 \leq a_0 \leq  8.43 \times 10^{-6} ~\text{m/s}^2.
\end{equation}
although, as~already mentioned before, in~the discussion to follow we shall use the tighter limits from~Mercury.

\section{Geodesic Equations and Effective~Potential\label{sec4}}

In this section, we will determine the ISCOs and OSCOs following the steps previously discussed in \cite{Boonserm:2019nqq}. First, we assume a fixed static, spherically symmetric gravitational background of the form
\begin{equation}
ds^2 = g_{tt} dt^2 - g_{rr} dr^2 - r^2 [ d \theta^2 + \sin^2\theta d \phi^2 ]
\end{equation}
 
Now, following  \cite{Garcia:2013zud}, the~equations of motion for test particles are given by 
\begin{equation}
\frac{d^2x^\mu}{ds^2} + \Gamma^\mu_{\rho \sigma} \frac{dx^\rho}{ds} \frac{dx^\sigma}{ds} = 0
\end{equation}
with $s$ as the proper time. The~corresponding Christoffel symbols, $\Gamma^\mu_{\rho \sigma}$, are computed by~\cite{landau}
\begin{equation}
\Gamma^\mu_{\rho \sigma} = \frac{1}{2} g^{\mu \lambda} \left( \frac{\partial g_{\lambda \rho}}{\partial x^\sigma} + \frac{\partial g_{\lambda \sigma}}{\partial x^\rho} - \frac{\partial g_{\rho \sigma}}{\partial x^\lambda} \right).
\end{equation}
 
The mathematical treatment is simplified taking advantage of the fact that there are two conserved quantities (two first integrals of motion), precisely as in the Keplerian problem in classical mechanics. In~practice,  for~$\mu=1=t$ and $\mu=4=\phi$, the~geodesic equations acquire the form
\begin{eqnarray}
0 & = & \frac{d}{ds} \left( g_{tt} \frac{dt}{ds} \right) \\
0 & = & \frac{d}{ds} \left( r^2 \frac{d\phi}{ds} \right).
\end{eqnarray}
 
With the above in mind, we then introduce the corresponding conserved quantities as
\begin{equation}
E \equiv g_{tt} \frac{dt}{ds}, \; \; \; \; \; \; L \equiv r^2 \frac{d\phi}{ds}.
\end{equation}
 
The last two quantities, $\{E, L\}$, are usually identified as the energy and angular momentum, respectively.
%
%
 
Assuming a motion on the $(x-y)$ plane (i.e., studying motions on the equatorial plane: $\theta = \pi/2$), the~geodesic equation for the $\theta$ index is also satisfied automatically. Therefore, the~only non-trivial equation is obtained for $\mu=2=r$ (see~\cite{Garcia:2013zud} for further details)
\begin{equation}
\left( \frac{dr}{ds} \right)^2 = \frac{1}{g_{tt} g_{rr}} \: \left[ E^2 - g_{tt} \left( \epsilon + \frac{L^2}{r^2} \right) \right]
\end{equation}
which may be also obtained from~\cite{Garcia:2013zud}
\begin{equation}
g_{\mu \nu} \frac{dx^\mu}{ds} \frac{dx^\nu}{ds} = \epsilon
\end{equation}
where $\epsilon = 1$ for massive test particles, and~$\epsilon = 0$ for light rays.
In the discussion to follow, we shall consider the case where $\epsilon = 1$ (massive test particle with mass $m$)
and $g_{tt} g_{rr} = 1$. Then, the non-trivial geodesic equation takes the simpler form
\begin{equation}
\left( \frac{dr}{ds} \right)^2 = \left[ E^2 - g_{tt} \left( 1 + \frac{L^2}{r^2} \right) \right]
\end{equation}
and we now introduce the corresponding effective potential, which, as usual, is defined to be
\begin{align}
V(r) &= g_{tt}(r) \left( 1 + \frac{L^2}{r^2} \right),
\end{align}
where $g_{tt}$ is the lapse function, which will now be identified to $A(r)$ reported in \mbox{Equation~(9)} (setting $\eta=0$).

\section{ISCOs/OSCOs in Massive~Gravity\label{sec5}}


From now on, we will investigate the case of massive particles (which means $\epsilon = 1$). The~effective potential in this case looks like
\begin{align}
V(r) &=  \bigg( 1 - \frac{2 M}{r} - \frac{1}{3} \Lambda  r^2
+ \gamma  r  \bigg) 
\bigg(
1 + \frac{L^2}{r^2}
\bigg)
\end{align}
and the first and second derivatives of the potential are
\begin{align}
&V'(r) =\frac{6 L^2 M}{r^4}  - \frac{2 L^2}{r^3} + \frac{2 M}{r^2}  -\frac{2 r}{l^2} +  \left(1-\frac{L^2}{r^2}\right)\gamma
\\
&V''(r) = -\frac{2}{l^2}-\frac{24 L^2 M}{r^5} + \frac{6 L^2}{r^4}-\frac{4 M}{r^3} + \frac{2 L^2}{r^3} \gamma
\end{align}
 
The circular orbits are obtained  demanding that
\begin{align}
&\dot{r} = 0,
\hspace{1cm} 
\text{and} 
\hspace{1cm}
\ddot{r}=0,
\end{align}
as function of the rest of parameters. The~latter means that we need to find the roots of $V'(r)$ for $r \equiv r(L,m,l,\gamma)$. 
In general, it is not possible to obtain an analytic solution, which is the present case. In~order to make progress, we can take an alternative route. From~$V'(r)=0$, we find $L^2$ and evaluate it on $V''(r)$ to find the $r_{\text{ISCO}}$ and $r_{\text{OSCO}}$. Thus, we have
\begin{align}
L^2 &= -\frac{r^2 \left(2 l^2 M + \gamma  l^2 r^2-2 r^3\right)}{l^2 \left(6 M-\gamma  r^2-2 r\right)}
\end{align}
 
Similarly to the Kottler spacetime, we reinforce that the angular momentum is real and finite for $r \in (r_{\text{ICCO}}, r_{\text{OCCO}}]$. Now, replacing $L^2$ into $V''(r)$, we have
\begin{align}
\begin{split}
V''(r) = &2\frac{-6 l^2 M+l^2 r-3 r^3}{l^2 r^3}+\frac{2\gamma }{r} -
\\
&
\frac{4 \left(24 l^2 M^2-10 l^2 M r+l^2 r^2-6 M r^3+r^4\right)}{l^2 r^3 \left(-6 M+\gamma  r^2+2 r\right)}
\end{split}
\end{align}
 
To obtain the corresponding roots of $V''(r)$, we have to solve the polynomial expression
\begin{align}
P_5(r) = b_5 r^5 + b_4 r^4 + b_3 r^3 + b_2 r^2 + b_1 r + b_0 
\end{align}
where the parameters are defined as 
\begin{align}
b_5 &=-3 \gamma
\\
b_4 &=\gamma ^2 l^2-8
\\
b_3 &=3 \gamma  l^2+30 M
\\
b_2 &=-12 \gamma  l^2 M
\\
b_1 &=2 l^2 M
\\
b_0 &=-12 l^2 M^2
\end{align}
where all lengths are expressed in parsec setting $l=5~\text{Gpc}$. There are five roots in total, which, in general, include real (positive or negative) as well as complex roots. We recall 
that, in the case of the Schwarzschild geometry, there is only one root, $r_{\text{ISCO}}=6 M$ \cite{Boonserm:2019nqq}. Given 
that there is no analytic expression for the roots of a fifth order polynomial, we shall compute the roots numerically once the numerical values of the parameters are~specified. 

Thus, we consider three numerical values of the massive gravity parameter $\gamma$ to exemplify how $r_{\text{ISCO}}$ and $r_{\text{OSCO}}$ vary for different structures in the Universe.
The first value of $\gamma$ is taken from~\cite{Panpanich:2018cxo}, $\gamma \sim 10^{-28}~\text{m}^{-1}$, whereas the remaining 
two values are obtained  in Section~\ref{sec3}. This is the second main result of the present work summarized in Table~\ref{numerical_table}.

Similarly to the Kottler spacetime, both ISCOs and OSCOs appear. Their numerical values are shown in Table~\ref{numerical_table}, while the astrophysical relevance is shown in Table~\ref{numerical_table_II}, considering typical values of the mass and size of known structures in the Universe. In~particular, our numerical results show that, in all cases, the ISCOs equal $6M$, which is precisely the Schwarzschild result. As~far as the OSCOs are concerned, in~two of the cases ($\gamma_1$ and $\gamma_2$), they do not depend on the mass of the astrophysical object. Despite the fact that the OSCOs obtained in those cases are not cosmologically large, their sizes ($2.89 \times 10^{7}~\text{pc}$ and $2.42 \times 10^{6}~\text{pc}$, respectively) are similar to the size of cluster of galaxies, which is very large compared to the dimensions of the astrophysical objects displayed in Table~\ref{numerical_table}.

In the third case ($\gamma_3$), the OSCOs computed here increase with the mass of the astrophysical object. In addition, their sizes are lower than the ones obtained in the Kottler spacetime~\cite{Boonserm:2019nqq}. In~this sense, the~OSCOs analysis within the framework of four-dimensional massive gravity reinforces their astrophysical importance. Finally, the~fact that the numerical values of the OSCOs obtained here are significantly different than the ones presented in~\cite{Boonserm:2019nqq} indicates that the $\gamma$ term is the dominant one, rather than the cosmological constant~term.

\section{Conclusions\label{sec6}}

In summary, we  studied the impact of a non-vanishing (positive) cosmological constant on the innermost and outermost stable circular orbits (ISCOs and OSCOs, respectively) within four-dimensional massive gravity. The~gravitational field generated by a point-like object is known, and,~at the non-relativistic limit, the gravitational potential differs by the Schwarzschild--de Sitter geometry by a term that is linear in the radial coordinate. The~numerical value of parameter $\gamma$ of the new, additional term may be determined either using data from the galaxy rotation curves or using data from the periastron advance in the solar system (planet Mercury) and in the Galactic center ($S_2$ star). 

Starting from the geodesic equations for massive test particles, and~the corresponding effective potential, we  obtained a polynomial of fifth order that allowed us to compute the innermost and outermost stable circular orbits. We  computed its roots numerically for several different structures in the Universe of increasing mass (from the hydrogen atom to stars and globular clusters to galaxies and galaxy clusters) considering three distinct values of the parameter $\gamma$, determined using physical considerations. 

Similarly to the Kottler spacetime, both ISCOs and OSCOs appeared. In~particular, our numerical results showed that the ISCOs equaled $6M$ (the Schwarzschild result) in all cases; whereas, for OSCOs, in~two of the cases ($\gamma_1$ and $\gamma_2$), this did not depend on the mass of the astrophysical object. In~spite of the fact that the OSCOSs obtained in those cases were not cosmologically large, their sizes ($2.89 \times 10^{7}$ and $2.42 \times 10^{6}$, respectively) were similar to the supercluster size, which is very large compared to the dimensions of the astrophysical objects displayed in Table~1. 

In~the third case ($\gamma_3$), the OSCOs obtained in the present work increased with the mass of the astrophysical object. In addition, their sizes were lower than those obtained in the Kottler spacetime. In~this sense, the~OSCOs analysis within the framework of four-dimensional massive gravity reinforces their astrophysical importance.

Finally, our numerical results indicate that, within massive gravity, the~parameter $\gamma$ played a crucial role in the determination of ISCOs and, more importantly, for OSCOs. Thus, it is $\gamma$, rather than $\Lambda$, as the term that mainly modifies the stable circular orbits, contrary to the Kottler spacetime, where $\Lambda$ is the term producing the new features as far as the OSCOs are~concerned.

\section*{Acknowlegements}

We are grateful to the anonymous reviewers for their constructive criticism as well as numerous useful comments and suggestions.
The authors \'A.R. and N.C. acknowledge Universidad de Santiago de Chile for financial support through the 
Proyecto POSTDOC-DICYT, C\'odigo 043131 CM-POSTDOC.
The authors G.P. and I.L. thank the Funda\c c\~ao para a Ci\^encia e Tecnologia (FCT), Portugal, 
for the financial support to the Center for Astrophysics and Gravitation-CENTRA, Instituto Superior T\'ecnico, 
Universidade de Lisboa, through the Project No.~UIDB/00099/2020 and grant No. PTDC/FIS-AST/28920/2017.


%


\end{document}